\documentclass[prc,superscriptaddress,unsortedaddress,twocolumn]{revtex4-2}
\usepackage[english]{babel}
\usepackage{graphicx,epsf,bm,amsmath,color}
\usepackage{here,ulem}
\usepackage{hyperref}
\usepackage{orcidlink}
\newif\iffigure
\figuretrue
\begin{document}
\title{Comparison of Relativistic and  Non-relativistic Faddeev Calculations 
for Proton-Deuteron Elastic Scattering}

\author{H. Kamada \orcidlink{0000-0001-6519-9645}} \email{kamada@rcnp.osaka-u.ac.jp}
\affiliation{Research Center for Nuclear Physics, Osaka University, Ibaraki 567-0047,
Japan}
\affiliation{Department of Physics, Kyushu Institute of Technology, 1-1 Sensuicho, Tobata, Kitakyushu 804-8550, Japan}

\author{A. Arslanaliev \orcidlink{0000-0002-8667-9688}}\email{arslanaliev.kh@gmail.com}
\author{Y. Kostylenko  \orcidlink{0000-0003-3869-2885}}\email{nlokost@gmail.com}
\author{A. V. Shebeko  \orcidlink{0000-0003-3357-2286}}\email{shebeko@kipt.kharkov.ua}
\affiliation{ Institute for Theoretical Physics NRC “Kharkiv Institute of Physics and Technology,” Kharkiv, Ukraine}
\author{J. Golak  \orcidlink{0000-0002-5210-6910}}\email{jacek.golak@uj.edu.pl}
\author{R. Skibi\'nski  \orcidlink{0000-0003-0806-4634}}\email{roman.skibinski@uj.edu.pl}
\author{K. Topolnicki  \orcidlink{0000-0002-9312-1842}}\email{kacper.topolnicki@uj.edu.pl}
\author{H. Wita\l a \orcidlink{0000-0001-5487-4035}}\email{henryk.witala@uj.edu.pl}
\affiliation{M. Smoluchowski Institute of Physics, Faculty of Physics, Astronomy and Applied Computer Science, Jagiellonian University, PL-30348 Krakow, Poland}
\author{V. Chahar \orcidlink{0009-0007-3310-8447}}\email{vaibhav.chahar@doctoral.uj.edu.pl}
\author{D. F. Ram\'irez Jim\'enez \orcidlink{0000-0002-2932-1399}}\email{df.ramirez@doctoral.uj.edu.pl}
\affiliation{ Jagiellonian University, Doctoral School of Exact and Natural Sciences, PL-30348 Krak\'ow, Poland } 
\affiliation{M. Smoluchowski Institute of Physics, Faculty of Physics, Astronomy and Applied Computer Science, Jagiellonian University, PL-30348 Krakow, Poland}
\author{W. N. Polyzou \orcidlink{0000-0001-9014-1250}}\email{polyzou@uiowa.edu}
\affiliation{Department of Physics and Astronomy, The University of Iowa, Iowa City, Iowa 52242, USA}

\begin{abstract}
This investigation compares non-relativistic and relativistic nucleon-nucleon 
potentials in the context of proton-deuteron scattering. 
Conventional NN potentials (e.g., CDBonn, AV18, Nijmegen) rely on the nonrelativistic Schr\"odinger equation, whereas the Kharkiv potential is intrinsically relativistic. We employ the Coester-Pieper-Serduke
(CPS) 
and Kamada-Gl\"ockle 
(KG) 
conversion methods to construct a phenomenological-relativistic potential 
(PRP) 
from a realistic NN potential, preserving the deuteron binding energy and phase shifts.  
Focusing on relativistic effects and not including Coulomb forces to avoid complexity, the solutions are compared by solving relativistic and nonrelativistic Faddeev equations.
Calculations of the differential cross section using the relativistic Faddeev equation show that relativistic effects—particularly the deviation at the backward angle—become pronounced at $135 \text{ MeV}$. The differences in the forward angle were attributed to the characteristics of the Kharkiv potential itself. The reverse transformation of the Kharkiv potential into a pseudo-nonrelativistic potential (PNRP) 
confirms that the backward-angle relativistic effect increases with energy in the range from $100 \text{ MeV}$ to $400 \text{ MeV}$. 
Comparisons of the polarization observables indicate that relativistic effects, as well as the discrepancy between the CPS and KG transformations, become significant above $300 \text{ MeV}$. 
However, for polarization observations below $300 \text{ MeV}$, the nonrelativistic results from PNRP do not deviate significantly from relativistic calculations.
\end{abstract}

\maketitle

\section{Introduction}

Fortunately, the existence of a bound state (deuteron) for the proton and neutron allows for the direct study of two-nucleon forces (2NF) and three-nucleon (3N) forces (3NF) by performing typical Nucleon-deuteron scattering experiments on a three-nucleon system \cite{Gloeckle96}. In the low-energy region, early studies \cite{Epelbaum2002,Entem2002} that provided initial insights have been followed by more quantitative analyses \cite{Reinert2018,Witala2022} based on modern chiral effective field theory ($\chi$EFT).

In contrast to conventional meson-theoretical potentials (CDBonn \cite{CDBonn}, AV18 \cite{AV18}, Nijmegen \cite{Nijmegen}), $\chi$EFT constructs a Lagrangian for the pion field (the Goldstone boson) and other background fields while respecting chiral symmetry. As the chiral order increases, it accurately reproduces the detailed structure of nuclear forces. A characteristic feature of this approach is that it derives potentials not only for 2N but also for 3N and many-nucleon forces from a single Lagrangian, with the coupling constants for 2NF and 3NF determined complementarily rather than independently as the chiral order increases.

Since $\chi$EFT is developed for the low-energy region, its applicability may be limited when considering a fully relativistic kinetic framework at higher energies. Nevertheless, many-body forces, such as two-body and three-body interactions, can be consistently constructed using an appropriate Lagrangian as a basis, not limited to $\chi$EFT. At high energies, a potential that strictly conforms to the kinematics of special relativity is required. 

The relativistic treatment given here means that if spin is considered to be kinematic, then we can use the Bakamjian-Thomas method to construct a dynamical unitary representation of the Poincaré group, with its mass operator as the dynamical mass operator.
Therefore, a framework for accurately handling Lorentz boost transformations in three-body systems has also been established \cite{Kamada2007,Coester1982,Kamada2002,Keister2006,Polyzou2019,Witala2011_1}.

We employ the nucleon–nucleon interaction derived using the unitary clothing transformation (UCT) \cite{Dubovik2010,Shebeko2012} based on the notion of clothed particles. This approach starts from a primary field-theoretical Lagrangian for interacting ($\pi-, \eta-, \delta-, \omega-, \rho-, \sigma-$) meson and nucleon fields, one comes to the corresponding Hamiltonian which interaction part contains relativistic interactions responsible for physical processes. The realization of relativistic invariance within the UCT method is discussed in detail in Ref. \cite{Shebeko2012_2}. In Ref. \cite{Dubovik2010}, the UCT method was used to derive the nucleon–nucleon interaction operator, known as the Kharkiv potential\cite{Kamada2017,Arslanaliev2021,Arslanaliev2022} that describes the NN$\to$NN scattering. The authors of \cite{Dubovik2010} do not consider the NN $\to$NN$\pi$ interaction operator that is justified at the NN collision energies below 300 MeV. Similar interactions appear during the clothing procedure (cf. decomposition (65) in \cite{KorCanShe2007}). These processes are not considered in the present work. 

Efforts to extend Faddeev three-body equations to the relativistic regime began with calculations of the triton binding energy \cite{Gloeckle1986,Kamada2002}. However, when using realistic potentials, a transformation from a non-relativistic potential to a relativistic one was required. Although there is no direct way to convert a non-relativistic potential (belonging to a different theoretical framework) into a relativistic potential, two approaches allow the binding energy and phase shifts to be preserved: one involves transforming the relativistic potential without changing the wave function, and the other involves a momentum scale transformation. These are referred to as the Coester-Pieper-Serduke (CPS) transformation \cite{Coester1975} and the Kamada-Gl\"ockle (KG) transformation \cite{Kamada1998}, respectively. 
The resulting forces are collectively referred to as phenomenological-relativistic potentials (PRPs).
While  the the CPS and KG transformations give exact dynamical representations of the Poincar\'e group \cite{Coester1965},  it is not known how to construct consistent
currents or three-body interactions.

Elastic electron-deuteron scattering in the one-photon-exchange approximation has been used to investigate the sensitivity of three-body observables to these transformations \cite{Allen2000}. Using realistic potentials such as CDBonn \cite{CDBonn}, PRPs were obtained via the KG transformation, and bound \cite{Kamada2002} and scattering states \cite{Witala2005} were analyzed in three-body systems. Because PRPs can be readily transformed and then subjected to Lorentz boost transformations \cite{Kamada2007}, the CPS transformation has been employed to calculate the $A_y$ polarization \cite{Witala2008} and three-body breakup reactions \cite{Witala2011_2,Polyzou2011}.

When using the Kharkiv potential, such transformations are unnecessary. However, because purely relativistic calculations are possible with the Kharkiv potential—and both the CPS and KG transformations are invertible—it is possible to verify the accuracy of previous non-relativistic Faddeev three-body calculations by effectively inverting the Kharkiv potential. 
In this sense, the inverse transformations can be viewed as a reduction of relativistic calculations to a non-relativistic framework.
Here, we are interested not only in comparing relativistic and non-relativistic calculations using the Kharkiv potential but also in analyzing differences between the CPS and KG transformations in three-body systems.

In the next section, we first compare the results of solving the relativistic Faddeev three-body equations using the relativistic Kharkiv potential with calculations employing conventional realistic potentials. Section \ref{Sec_III} provides a brief introduction to the CPS and KG transformations, followed by a comparison between them in Sec. \ref{Sec_IV}. The summary is given in Sec. \ref{Sec_V}.

In all Faddeev calculations presented in this paper, the total three-body angular momentum was taken from $J=\tfrac{1}{2}$ to $\tfrac{25}{2}$ with both parities $\pm$, and the two-body subsystem angular momentum from $j=0$ to 5. We verified that the numerical results were sufficiently converged with respect to the partial-wave expansion.

\section{Kharkiv Potential}
\label{Sec_II}
As mentioned above, the Kharkiv potential can be inserted directly into the relativistic Faddeev equations without the need for a transformation to PRP.
First, we present the differential cross sections for proton–deuteron elastic scattering at low energies, comparing realistic potentials (CDBonn \cite{CDBonn}, AV18 \cite{AV18}, Nijmegen \cite{Nijmegen}) with the Kharkiv potential.
 In recent years, we have succeeded in incorporating Coulomb forces into non-relativistic Faddeev calculations \cite{Witala2024}, but we have not yet been able to directly integrate them into our relativistic Faddeev calculations.
Since the effect of the Coulomb force becomes negligible when the scattering angle exceeds about 20 degrees, except in the low energy region, and because we are focusing on relativistic effects, we ignore the Coulomb force in our calculations.

\begin{figure}[h]
\centering
\includegraphics[width=0.45\textwidth]{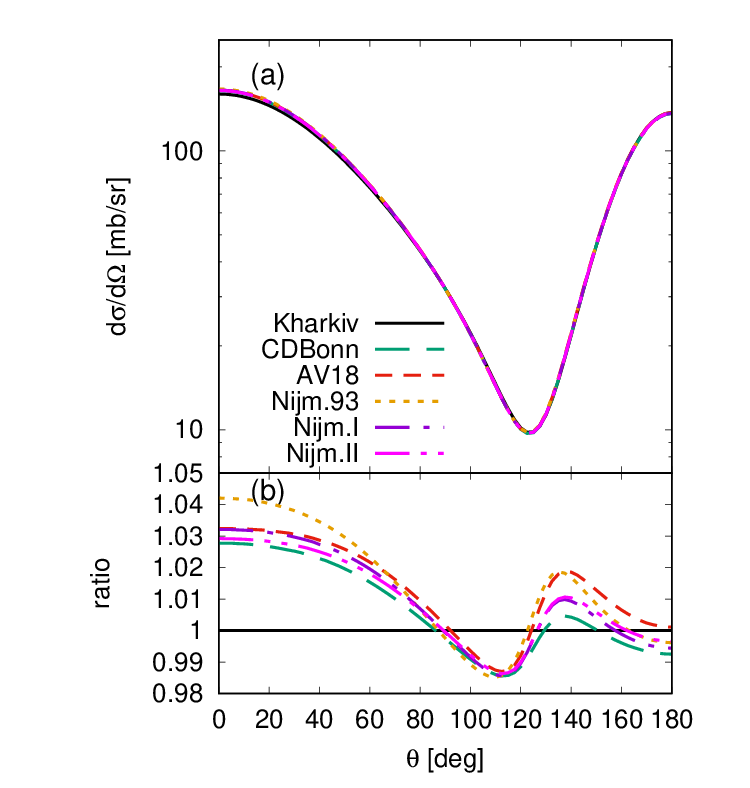}
\caption{
Differential cross section of p-d elastic scattering
at an incident proton laboratory energy of E= 13 MeV.
(a) Results of non-relativistic calculations performed using the CDBonn \cite{CDBonn} 
(long-dashed green line), AV18 \cite{AV18} (short-dashed red line), 
Nijmegen 93 \cite{Nijmegen} (dotted orange line), Nijmegen I (single-dot–dashed violet line), 
and Nijmegen II (double-dot–dashed magenta line) potentials are shown, 
together with the relativistic calculation based on the Kharkiv potential \cite{Arslanaliev2022}
(solid black line).
(b) Relative differences of the non-relativistic results shown in panel (a) 
with respect to the predictions based on the Kharkiv potential for the differential cross section.
}
\label{fig:1}
\end{figure}

\begin{figure}[h]
\centering
\includegraphics[width=0.45\textwidth]{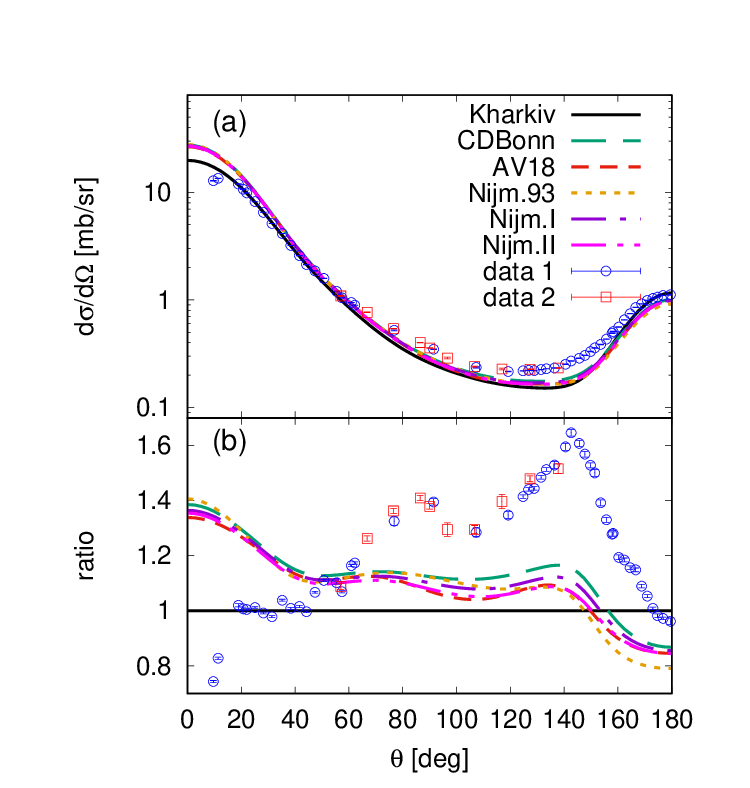}
\caption{ 
Differential cross section of p-d elastic scattering at $E=$ 135 MeV.
The line colors are the same as in Fig. 1. Data sets 1 and 2 are taken from Refs. \cite{Sakai2000} and \cite{Sakamoto1996}, 
respectively.
}
\label{fig:2}
\end{figure}

The differential cross sections for low-energy and intermediate-energy proton–deuteron elastic scattering are shown in Fig. \ref{fig:1} and Fig. \ref{fig:2} for projectile kinetic energies of 13 MeV and 135 MeV, respectively.
At low energies, the theoretical values for all potentials follow nearly identical curves.
In other words, the expected relativistic effect is shown to be less than 5\% in this energy range, and this difference is considered unobservable given the experimental precision. See detail in Fig. \ref{fig:1}(b).

At low energies, the overall theoretical curves agree, but at intermediate energies, differences emerge between the forward and backward scattering angles.
At intermediate scattering angles, the discrepancy \cite{Koike1998} from the experimental values becomes large, but this can be explained \cite{Witala2001} by including a three-body force in addition to the conventional realistic meson-exchange potentials.


\begin{figure}[h]
\centering
\includegraphics[width=0.45\textwidth]{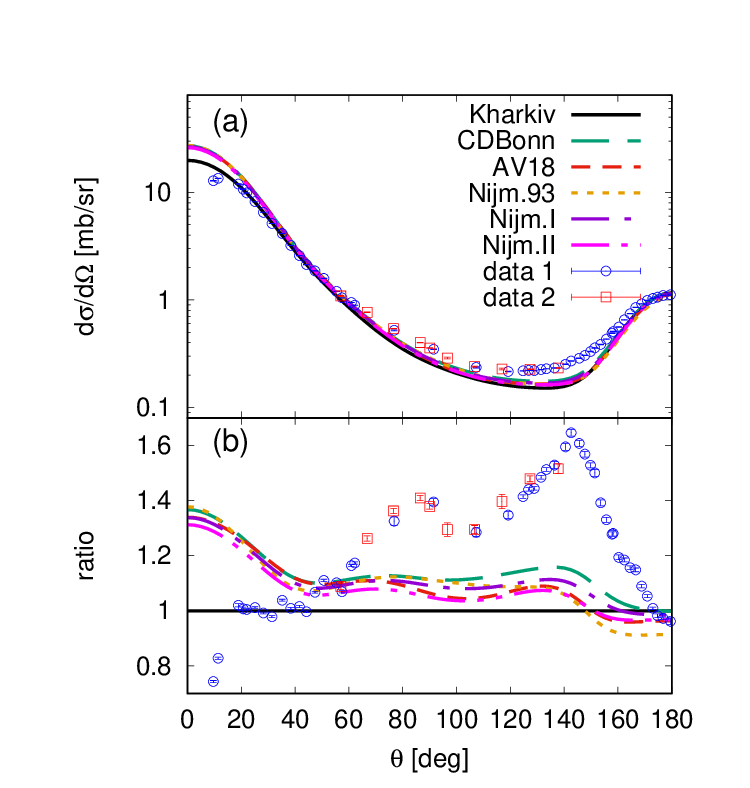}
\caption{
Differential cross section of p-d elastic scattering at $E=$ 135 MeV.
The meaning of each line is the same as in Fig. \ref{fig:2}. The realistic potentials 
(CDBonn, AV18, Nijmegen) are transformed into phenomenological 
relativistic 
potentials (PRP) using the CPS method and then employed in the relativistic 
Faddeev equations. The data and line types are the same as in Fig. \ref{fig:2}.
}
\label{fig:4}
\end{figure}

It is not straightforward to determine whether this difference is due to the properties of the potential or to the relativistic versus non-relativistic nature of the solution.
These realistic potentials are transformed into PRPs via CSP method and in the next step their predictions are compared. 
Figure \ref{fig:4} shows a comparison of the realistic PRPs with the Kharkiv potential at 135 MeV.
Apparently, in backward-angle scattering, the difference between the cross sections obtained with the Kharkiv potential and those obtained with other PRPs disappears.
The difference persists in the forward scattering region. To determine whether this discrepancy is specific to the Kharkiv potential, calculations must be performed that accurately include the Coulomb force. 
 As mentioned earlier, this intermediate energy calculation does not include the Coulomb force.
However, since the influence of the Coulomb force becomes insignificant for scattering angles above about 20 degrees, it is likely that the observed difference arises from the analysis method applied at the relativistic level of the Kharkiv potential.

\section{Non-relativistic reductions}
\label{Sec_III}
In the present context, the term non-relativistic reduction refers to the following construction: we begin with a potential $V$ that satisfies the relativistic Lippmann–Schwinger equation and reproduces the same phase shifts and deuteron binding energy as the original relativistic interaction. We then define a potential $V_{nr}$ that satisfies the non-relativistic Lippmann–Schwinger equation while preserving these observables. Because the relativistic and non-relativistic frameworks are based on different mechanical principles, this procedure is not a systematic reduction scheme in the traditional sense, and the non-relativistic treatment does not result in a relativistic approximation, and its accuracy cannot be arbitrarily improved in principle.
As emphasized above, this somewhat cautious terminology is intentional. The construction employed here should be regarded not as a reduction in the perturbative sense but rather as a definition: the procedure effectively reverses the standard construction of a phenomenological 
relativistic potential (PRP).
As discussed in the Introduction, two well-established methods are available to generate a PRP: the CPS transformation \cite{Coester1975, Coester1982} and the KG transformation \cite{Kamada1998}. For completeness, we summarize the essential features of both approaches in the following subsections.

\subsection{CPS transformation}
\label{subsec_CPS}

Within the CPS framework, a relativistic potential $V$ is mapped onto a non-relativistic potential $V^{CPS}_{nr}$.
The non-relativistic Schr\"odinger equation for two identical particles reads
\begin{eqnarray}
&&
\left ( {p^2 \over m } + V^{CPS}_{nr} \right) \Psi={p_0^2\over m} \Psi,
\label{S_nonrel}
\end{eqnarray}
where $m$ denotes the particle mass, $p$ the relative momentum, and $p_0$ the momentum eigenvalue.
The corresponding relativistic equation is
\begin{eqnarray}
&& \left( 2\sqrt{m^2+p^2} + V \right) \Psi = 2\sqrt{m^2+p_0^2}\Psi.
\label{S_rel}
\end{eqnarray}
The wave function $\Psi$ is identical in Eqs.~(\ref{S_nonrel}) and (\ref{S_rel}). Squaring the operator in Eq.~(\ref{S_rel}), one obtains
\begin{eqnarray}
&&
\left( 2\sqrt{m^2+p^2} + V \right)^2\Psi = 4(m^2+p_0^2) \Psi.
\end{eqnarray}
Eliminating $p_0$ via Eq.~(\ref{S_nonrel}) and solving for $V^{CPS}_{nr}$—noting that $p$ and $V_{nr}$ are operators—yields
\begin{eqnarray}
V^{CPS}_{nr}= {1\over 4 m} \left(2\sqrt{m^2+p^2}\, V + 2V\sqrt{m^2+p^2} +V^2 \right).
\end{eqnarray}
The corresponding partial-wave momentum-space representation is given explicitly in Ref.~\cite{Kamada2007} as
\begin{eqnarray}
&& V^{CPS}_{nr}(p,p')={1 \over 2m} \left( \sqrt{m^2+p^2} +\sqrt{m^2+p'^2} \right) V(p,p')
\cr &&\qquad + {1\over 4m } \int_0^\infty V(p,p'')V(p'',p') \, {p''}^2 \, dp''.
\end{eqnarray}

\subsection{KG transformation}
\label{KG_}

The KG transformation constitutes a momentum-scale transformation that ensures phase equivalence between relativistic and non-relativistic descriptions \cite{Kamada1998}.
For a system of two identical particles, we introduce the \textit{relativistic} relative momentum $p$ and the \textit{non-relativistic} momentum $k_{nr}$.
In the center-of-mass frame, the relativistic kinetic energy is
\begin{equation}
E=2\sqrt{m^2+p^2}-2m.
\end{equation}
The corresponding non-relativistic kinetic energy is
\begin{equation}
E_{\rm nr}=\frac{k_{nr}^2}{m}.
\end{equation}
The well-known expansion
\begin{equation}
2\sqrt{m^2+p^2}-2m=\frac{p^2}{m} + \cdots
\end{equation}
demonstrates that $E_{\rm nr}$ approximates $E$ when $k_{nr}$ is identified with $p$.
In the KG construction, however, $k_{nr}$ is \textbf{defined} by enforcing the identity
\begin{equation}
E=2\sqrt{m^2+p^2}-2m \stackrel{\mathrm{def}}{=} \frac{k_{nr}^2}{m}= E_{nr},
\end{equation}
implying $k_{nr}\ne p$ in general. Thus, $p$ determines $k_{nr}$ and, conversely, $k_{nr}$ determines $p$.

The non-relativistic Schr\"odinger equation,
\begin{equation}
\left( \frac{k_{nr}^2}{m} +V^{KG}_{nr} \right) \psi_{nr} = E_{nr} \psi_{nr},
\end{equation}
is therefore equivalent to the relativistic Schr\"odinger equation,
\begin{equation}
\left( 2\sqrt{m^2+p^2}-2m +V \right) \Psi = E \Psi,
\end{equation}
where $V^{KG}_{nr}$ and $V$ are the non-relativistic and relativistic potentials, respectively.

Remarkably, this mapping guarantees that the bound-state energy and scattering phase shifts are identical in both descriptions. The explicit relations among $k_{nr}$, $p$, the potentials, and the wave functions are \cite{Kamada1998}
\begin{eqnarray}
&& k_{nr} \stackrel{\mathrm{def}}{=} \sqrt{ 2m\left( \sqrt{m^2 +p^2} -m \right) },\cr
&& V^{KG}_{nr}(k_{nr},k_{nr}') \stackrel{\mathrm{def}}{=} h(k_{nr}) V(p(k_{nr}),p'(k_{nr}')) h(k_{nr}') , \cr
&& \psi_{nr} (k_{nr})\stackrel{\mathrm{def}}{=} \Psi (p(k_{nr})) h(k_{nr}),
\end{eqnarray}
with
\begin{eqnarray}
&& h(k_{nr})=\sqrt{\left(1+\frac{k_{nr}^2}{2m^2} \right)\sqrt{1+\frac{k_{nr}^2}{4m^2}}}.
\end{eqnarray}

\section{Comparison of the CPS and KG transformations}\label{Sec_IV}

In Sec.~\ref{Sec_II}, Fig. \ref{fig:4} 
we compared the solutions of the relativistic Faddeev three-body equation obtained using the PRPs constructed from a non-relativistic realistic potential with those obtained using the fully relativistic Kharkiv potential. In this section, we discuss the differences between the CPS and KG transformations by applying both inverse transformations to the Kharkiv potential and solving the resulting potentials within the non-relativistic Faddeev framework.

\begin{figure}[t]
\centering
\includegraphics[width=0.45\textwidth]{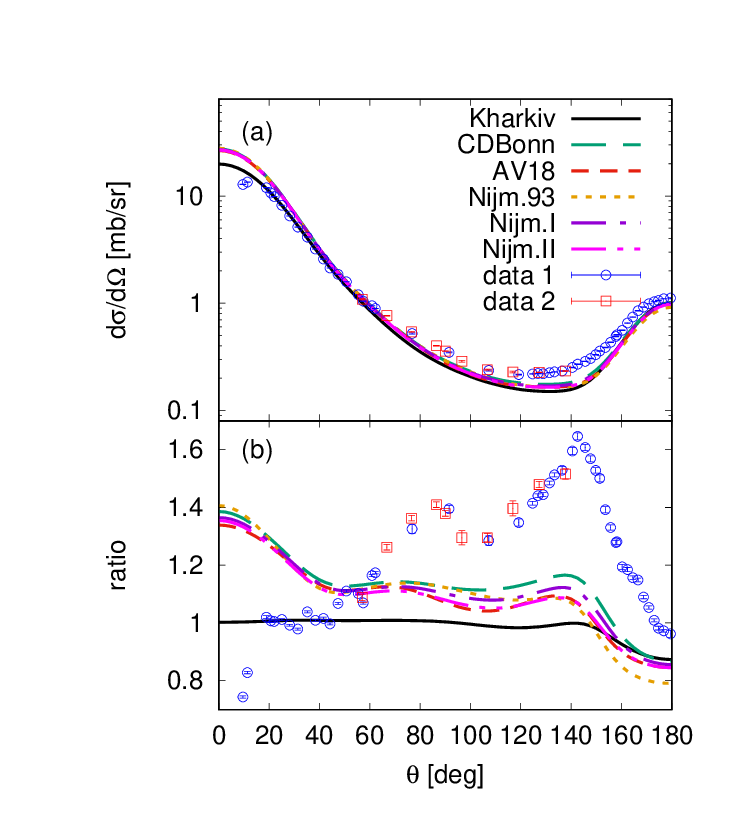}
\caption{
Differential cross section of p-d elastic scattering at $E= 135$ MeV.
The meaning of each line is the same as in Fig. \ref{fig:2}, except that the Kharkiv 
potential has been replaced by its PNRP obtained via the inverse CPS 
transformation,  and it is substituted into the non-relativistic Faddeev equations. The experimental data are the same as in Fig. \ref{fig:2}.
}
\label{fig:5}
\end{figure}

Let us consider the inverse procedure used to construct PRPs.
Figure~\ref{fig:5} shows the Kharkiv potential transformed into a pseudo-non-relativistic potential (PNRP) using the inverse CPS transformation, compared with standard realistic NN potentials (AV18 \cite{AV18}, CD-Bonn \cite{CDBonn}, and Nijmegen \cite{Nijmegen}) that have not been mapped into PRPs.

As discussed in Sec. \ref{Sec_II}, the overall consistency between the relativistic Faddeev calculations—especially at backward scattering angles—and their agreement with non-relativistic Faddeev results suggests that the primary dynamical features are robust under the transformation.
A similar level of agreement is observed at forward angles, independent of whether the calculation is performed relativistically or non-relativistically.

\begin{figure}[h]
\centering
\includegraphics[width=0.45\textwidth]{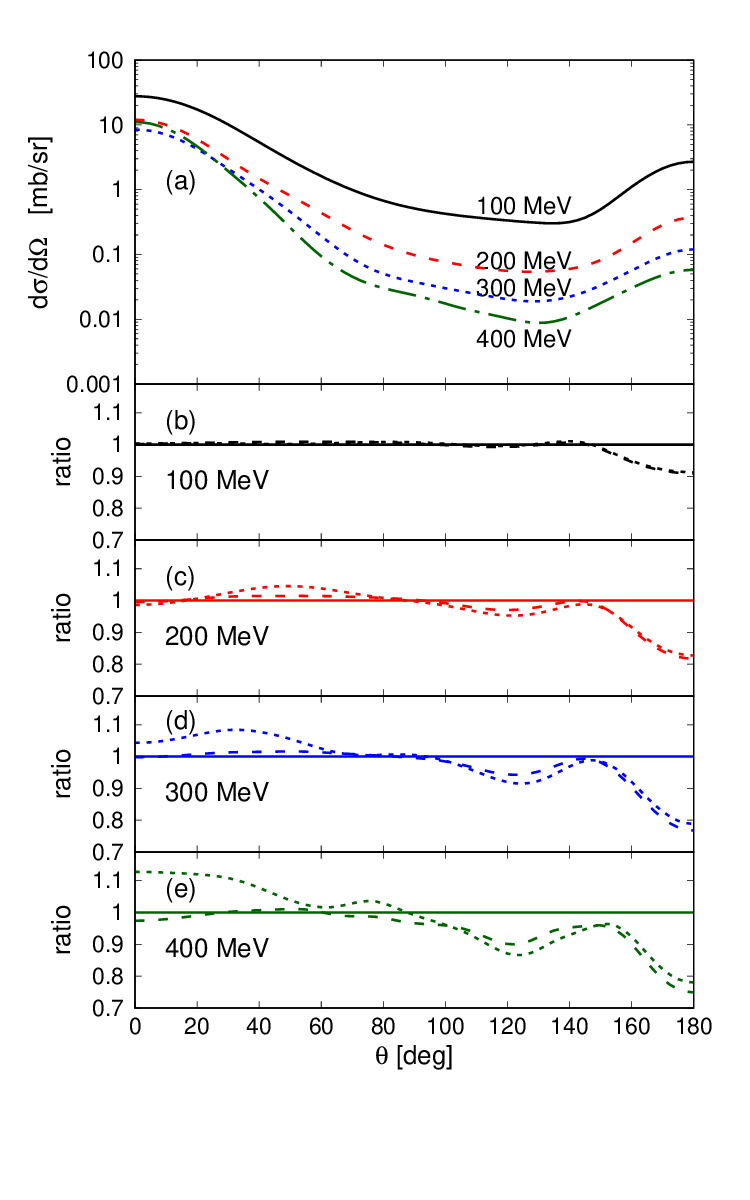}
\caption{
Differential cross sections calculated using the Kharkiv potential. 
(a) The calculation of the relativistic 
differential cross section for various incident proton 
energies.
The lines representing proton incidence energies of 100 MeV, 200 MeV, 300 MeV, and 400 MeV are depicted as solid black line, dashed red line, dotted blue line, and single-dotted green line, respectively.
(b) to (e) The relative differences (ratio vs. (a) ) of the non-relativistic 
calculation using CPS (dashed line) and KG (dotted line) transformation
with respect to the relativistic calculation at the energies considred in panel (a).
}
\label{fig:6}
\end{figure}

Figure~\ref{fig:6} presents the differential cross sections calculated relativistically with the original Kharkiv potential for incident energies of 100, 200, 300, and 400~MeV (solid black lines).
The corresponding non-relativistic results obtained using the CPS- and KG-based PNRPs are shown by the solid blue and dotted red lines, respectively.
As discussed in Sec.~\ref{Sec_II}, relativistic effects are most clearly visible at backward angles, a trend that is preserved in the PNRP calculations.
Within the examined energy range (up to 400~MeV), the differences between the CPS and KG transformations are small.

\begin{figure}[h]
\centering
\includegraphics[width=0.45\textwidth]{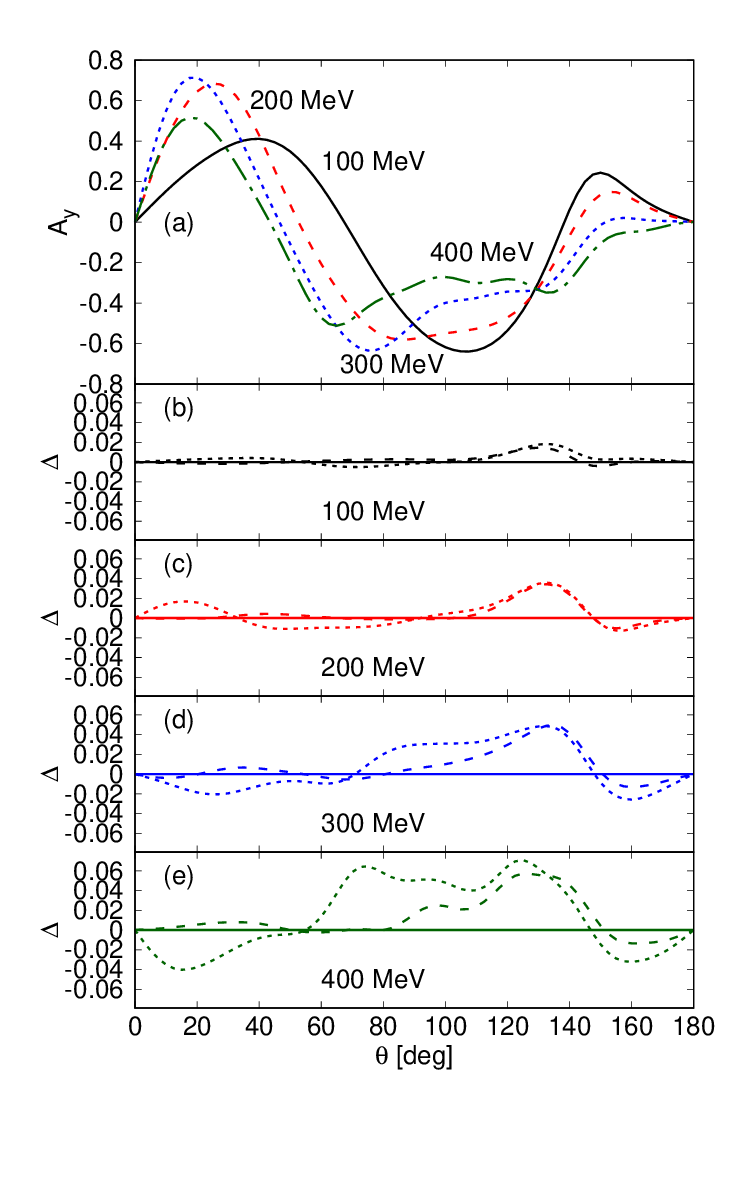}
\caption{
The same as in Fig. \ref{fig:6} but for the proton vector analyzing power Ay. 
Now in panels (b) to (e) just differences $ \Delta$ between non-relativistic (CPS and KG based) results 
and relativistic calculation are shown.
}
\label{fig:7}
\end{figure}

\begin{figure}[h]
\centering
\includegraphics[width=0.45\textwidth]{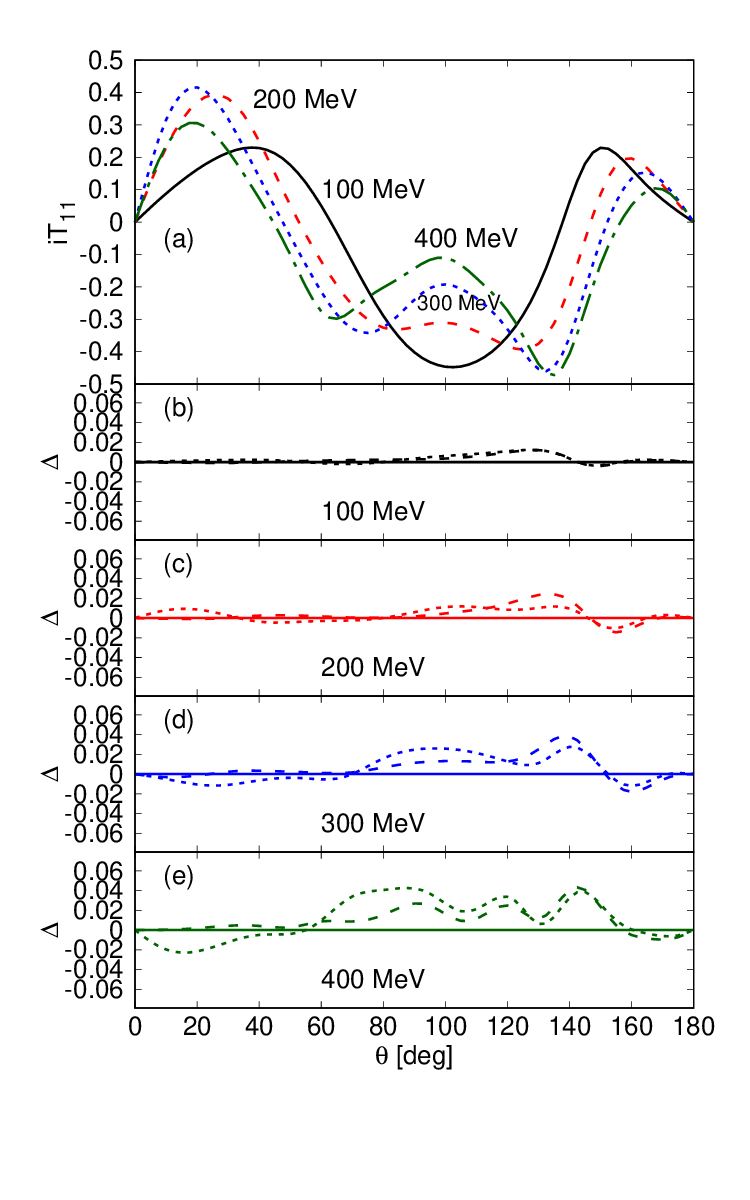}
\caption{
The same as in Fig. \ref{fig:7} but for the deuteron vector polarization $iT_{11}$. 
}
\label{fig:8}
\end{figure}

\begin{figure}[h]
\centering
\includegraphics[width=0.45\textwidth]{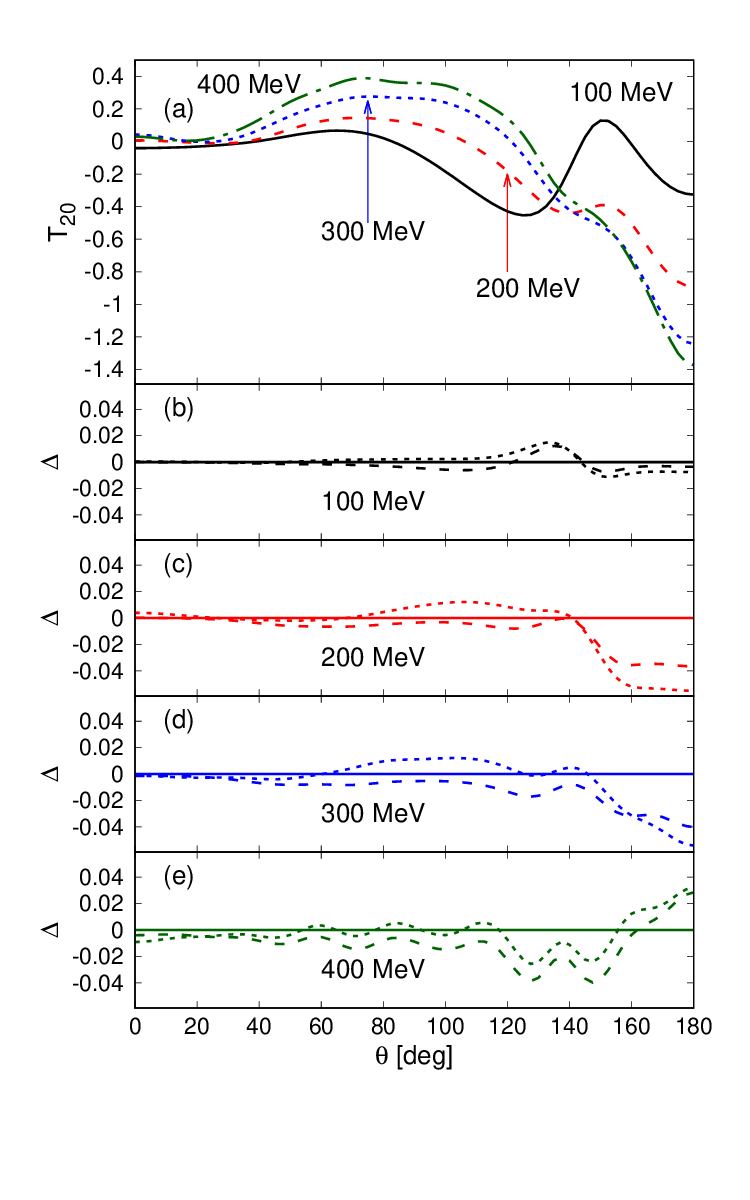}
\caption{ The same as in Fig. \ref{fig:7} but for the deuteron tensor polarization $T_{20}$.
}
\label{fig:9}
\end{figure}

\begin{figure}[h]
\centering
\includegraphics[width=0.45\textwidth]{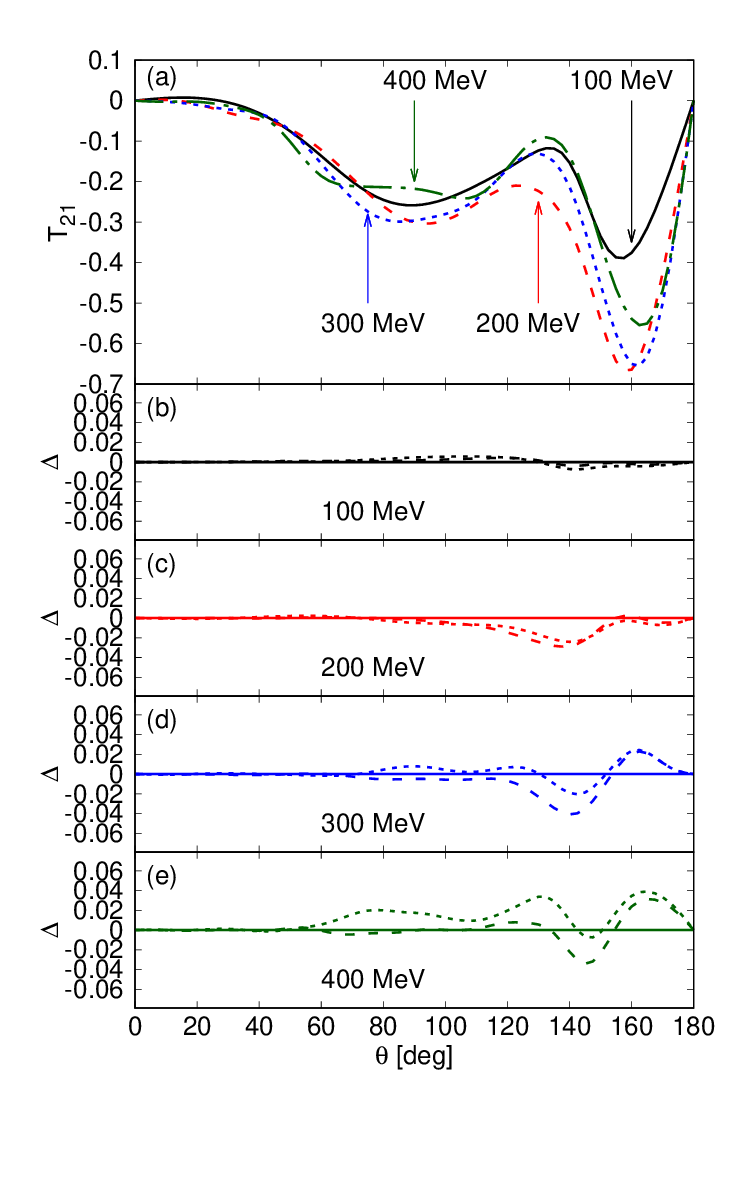}
\caption{ The same as in Fig. \ref{fig:7} but for the deuteron tensor polarization $T_{21}$.
}
\label{fig:10}
\end{figure}

\begin{figure}[h]
\centering
\includegraphics[width=0.45\textwidth]{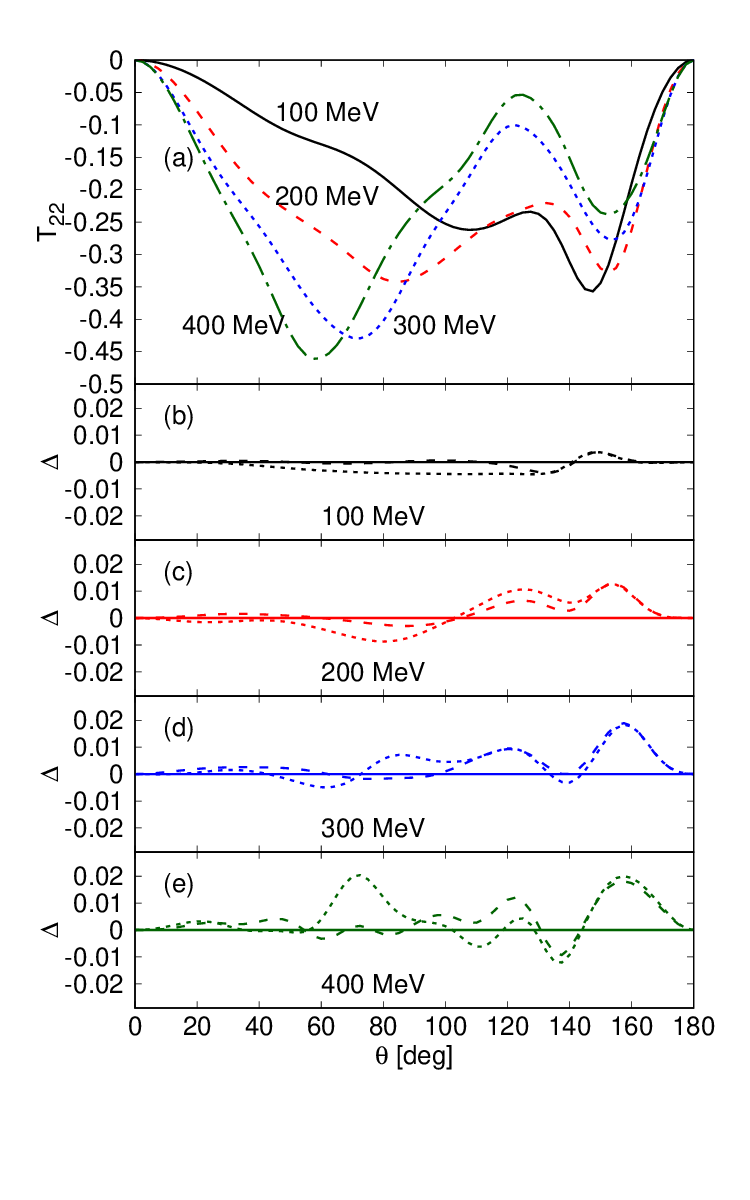}
\caption{ The same as in Fig. \ref{fig:7} but for the deuteron tensor polarization $T_{22}$.
}
\label{fig:11}
\end{figure}

To further assess differences between the transformations, we also examine typical polarization observables: the proton vector analyzing power $A_y$, the deuteron vector polarization $iT_{11}$, and the deuteron tensor components $T_{20}$, $T_{21}$, and $T_{22}$.
Figures~\ref{fig:7}–\ref{fig:11} present these observables under the same conditions as in Fig.~\ref{fig:6}.
Overall, the polarization values for both CPS and KG are in good agreement at forward angles, except for $iT_{11}$ and $A_y$. 
Figures \ref{fig:6}-\ref{fig:11} (b)-(e) show the differences between non-relativistic and relativistic 
observables for each energy, with the relativistic value set to zero. 
If the non-relativistic lines (based on KG or CPS transformations) are plotted together 
with the relativistic lines in panel (a), they are so close that it is difficult 
to distinguish between them.
However, as the angle becomes more backward, deviations from the relativistically accurate results become apparent, with CPS being closer to the accurate results than KG. Conversely, for $T_{22}$ at E=400MeV, KG sometimes comes closer to the accurate results at scattering angles around 120 degrees.

The relativistic Faddeev calculations shown here do not include the Wigner spin rotation.
As demonstrated in Ref.~\cite{Witala2005}, the impact of Wigner rotations on polarization observables is negligible below about 300~MeV.
At higher energies (above roughly 500~MeV), however, additional care is required, particularly because the NN database used to construct the Kharkiv potential extends only up to about 350~MeV in the laboratory frame.

\section{Summary}\label{Sec_V}

The conventional NN potentials (CD-Bonn, AV18, Nijmegen) are parameterized and constructed within the framework of the non-relativistic Schr\"odinger (or Lippmann–Schwinger) equation, whereas the Kharkiv potential is constructed in a fully relativistic manner. The CPS and KG transformations discussed in Sec.~\ref{Sec_III} serve as phase-preserving mappings that also maintain the correct deuteron binding energy. Using the CPS method, a realistic PRP was generated 
from conventional forces and 
inpremented in the relativistic Faddeev equation to compute the proton–deuteron elastic differential cross section, which was subsequently compared with the results obtained from the direct relativistic Kharkiv potential.

For an incident proton energy of 135~MeV, the cross section at backward scattering angles exhibits a clear relativistic effect, while differences at forward angles appear to be characteristic of the Kharkiv interaction. The CPS and KG transformations can also be inverted: applying them to the relativistic Kharkiv potential produces a corresponding pseudo-non-relativistic potential (PNRP), enabling systematic comparison with the full relativistic results. These comparisons show that the relativistic enhancement of the backward-angle cross section becomes increasingly pronounced as the energy rises from 100 to 400~MeV. The analysis of polarization observables demonstrates that relativistic effects become noticeable above approximately 300~MeV. The differences between the CPS and KG transformations also increase with energy, becoming significant above roughly 300~MeV.
Such a shift is expected due to differences in whether the non-relativistic phase
shifts are identified with the relativistic (experimental) phase shifts as functions of c.m. energy or c.m. momentum.

Nevertheless, the polarization observables in proton–deuteron elastic scattering remain reasonably well reproduced by 
calculations employing PRPs, provided that the incident energy does not exceed about 300~MeV.
Future work will include calculations of the Kharkiv three-nucleon force within the UCT framework.

\vspace{2.0cm}

{\it Acknowledgments.}
This work was partly supported by the Japan Society for the Promotion of Science (JSPS) under Grant No.~JP25K07301. Additional support was provided by the National Science Centre, Poland, under Grant No.~IMPRESS-U 2024/06/Y/ST2/00135, and by the Excellence Initiative–Research University Program at the Jagiellonian University in Krak\'ow. 
The Ukrainian team was also supported under the IMPRESS-U program by the National Academy of Sciences (USA) and the Office of Naval Research Global (USA), in assistance of the Science and Technology Center in Ukraine (grant No. 7134).
The numerical calculations were performed in part on the interactive server at RCNP, Osaka University, Japan, and on the supercomputers of the J\"ulich Supercomputing Center (JSC), J\"ulich, Germany.

\end{document}